\documentclass[sigconf]{acmart}

\AtBeginDocument{%
  \providecommand\BibTeX{{%
    \normalfont B\kern-0.5em{\scshape i\kern-0.25em b}\kern-0.8em\TeX}}}

\usepackage{stfloats}
\usepackage{graphics}
\usepackage{url}
\usepackage{hyperref}

\newcommand{\projectname}{GazeReader}

\copyrightyear{2023}
\acmYear{2023}
\setcopyright{rightsretained}
\acmConference[CHI EA '23]{Extended Abstracts of the 2023 CHI Conference on Human Factors in Computing Systems}{April 23--28, 2023}{Hamburg, Germany}
\acmBooktitle{Extended Abstracts of the 2023 CHI Conference on Human Factors in Computing Systems (CHI EA '23), April 23--28, 2023, Hamburg, Germany}\acmDOI{10.1145/3544549.3585790}
\acmISBN{978-1-4503-9422-2/23/04}

\begin{document}

\title[\projectname{}: Detecting Unknown Word Using Webcam ...]{\projectname{}: Detecting Unknown Word Using Webcam for English as a Second Language (ESL) Learners}


\author{Jiexin Ding}
\authornote{The authors contribute equally to this paper.}
\email{jxding17@gmail.com}
\affiliation{%
  \department{Global Innovation Exchange}
  \institution{Tsinghua University}
  \city{Beijing}
  \country{China}
}

\author{Bowen Zhao}
\authornotemark[1]
\email{bowen98@uw.edu}
\affiliation{%
    \department{Global Innovation Exchange}
  \institution{University of Washington}
  \city{Seattle}
  \country{United States}
}
\affiliation{%
  \institution{Tsinghua University}
  \city{Beijing}
  \country{China}
}

\author{Yuqi Huang}
\authornotemark[1]
\email{huangyuqi1999@gmail.com}
\affiliation{%
    \department{Global Innovation Exchange}
  \institution{University of Washington}
  \city{Seattle}
  \country{United States}
}
\affiliation{%
  \institution{Tsinghua University}
  \city{Beijing}
  \country{China}
}

\author{Yuntao Wang}
\email{yuntaowang@tsinghua.edu.cn}
\affiliation{%
    \department{Department of Computer Science and Technology}
  \institution{Tsinghua University}
  \city{Beijing}
  \country{China}
}
\authornote{denotes as the corresponding author.}

\author{Yuanchun Shi}
\email{shiyc@tsinghua.edu.cn}
\affiliation{%
  \department{Department of Computer Science and Technology}
  \institution{Tsinghua University}
  \city{Beijing}
  \postcode{100084}
  \country{China}
}

\renewcommand{\shortauthors}{Jiexin Ding, et al.}

\begin{abstract}
Automatic unknown word detection techniques can enable new applications for assisting English as a Second Language (ESL) learners, thus improving their reading experiences. 
However, most modern unknown word detection methods require dedicated eye-tracking devices with high precision that are not easily accessible to end-users. 
In this work, we propose \projectname{}, an unknown word detection method only using a webcam. 
\projectname{} tracks the learner's gaze and then applies a transformer-based machine learning model that encodes the text information to locate the unknown word. 
We applied knowledge enhancement including term frequency, part of speech, and named entity recognition to improve the performance. 
The user study indicates that the accuracy and F1-score of our method were 98.09\% and 75.73\%, respectively. Lastly, we explored the design scope for ESL reading and discussed the findings. 

  
\end{abstract}







\maketitle

\section{Introduction}
\label{sec:intro}

\begin{figure*}
  \centering
  \includegraphics[scale = 0.65]{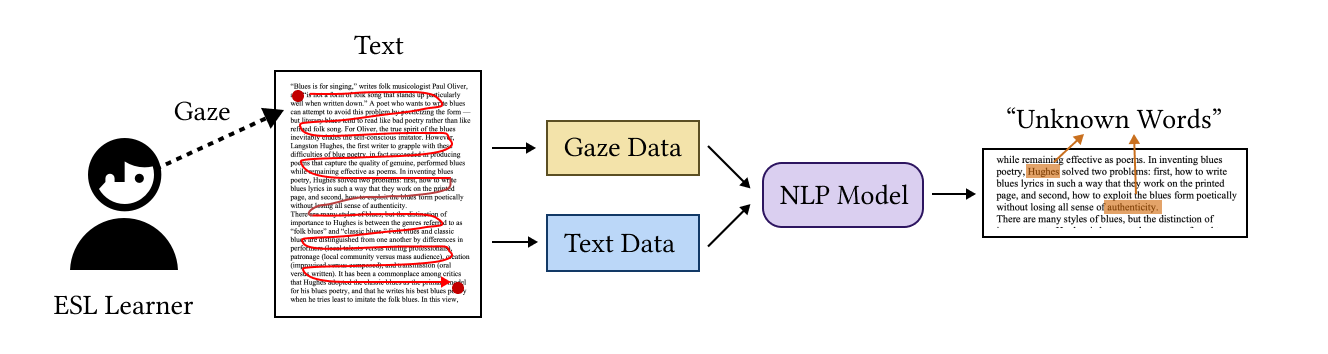}
  \caption{\projectname{} collects gaze data and text information and uses a transformer-based machine learning model to detect unknown words for ESL.}
  \label{fig:pipeline} 
\end{figure*}

Unknown words, which we defined as words that the user spends extra time thinking about, can significantly increase the reading difficulty for English as a Second Language (ESL) learners~\cite{rigg1991whole,mokhtar2012guessing}, since vocabulary knowledge is considered an essential feature of reading ability. Automatic detection of unfamiliar words can help users improve reading fluency and reading experience~\cite{hyrskykari2006eyes}. Previous works used eye movement features such as fixation duration and the number of regressions to detect unknown words, considering the strong correlation between eye movement and reading difficulty~\cite{unknown-word_hiraoka_2016, gaze-text_garain_2017, difficulties_lunte_2020}. However, all these methods are based on dedicated eye-tracking devices. Even for portable eye trackers, prices of hundreds or thousands of dollars prevent most people from using these technologies in their daily lives. 

The ubiquity of webcams makes it possible to track eye movement non-obtrusively. Researchers have used webcam-based eye-tracking methods to analyze users' reading behavior~\cite{eval-webcam_hutt_2022}, but the low precision~\cite{searchGazer_papoutsaki_2017} makes it infeasible to extract eye movement features proposed by previous works. Therefore, these reading analysis approaches cannot be directly transferred to webcam-based unknown word detection, which requires fine-grained tracking.

In this work, we detect unknown words using the webcam for ESL learners by integrating gaze and text information with the help of transformer-based language models. Our method tracks eye movement using WebGazer~\cite{papoutsaki2016webgazer} and embeds the positional information of gaze and texts with Long Short-term Memory~\cite{hochreiter1997long} (LSTM) models. In order to improve the model's performance given only noisy gaze data acquired by webcam, we leverage pre-trained language models to encode the context information for assistance. The gaze and context data are then fed to the model's classifier to predict the targeted unknown word. The F1-score of our model is 75.73\% which is higher than the text- and gaze-only model. We also conduct a user study to discover the needs of ESL learners and explore the design scope according to our method. The results show that our future design should focus more on proper nouns, multi-meaning words, and long and complex sentences, as well as exploring ways to reduce disturbance to the user's reading process through interactive design. In summary, the main contributions of this work include the following:

1) We propose \projectname{}, a novel webcam-based unknown word detecting method that leverages gaze and text information for ESL learners. The accuracy is 98.09\%,  the F1-score is 75.73\%, and the cross-user F1-score is 78.26\%.

2) We explore the design scope for ESL reading based on our method, guiding our future design direction towards proper nouns, multi-meaning words, and long and complex sentences.

\section{Background and Related work}
\label{sec:related}
It is time-consuming for second language learners to find the meaning of unknown words by looking up dictionaries or using translation tools. Automatically detecting unknown words can enable a variety of ways to facilitate reading. Researchers have implemented automatic unknown word detection by tracking users' clicks on a web document~\cite{web_ehara_2010} or combining text features with motion data on a smartphone~\cite{imu_higa_2022}. In addition, other works used eye movements to detect unknown words. The physiological and psychological reason behind it is that eye movements largely reflect people's tendency to pay attention when reading~\cite{rayner1998eye}. Word length, word frequency, and familiarity with the word largely explain the length of time users spend looking at it, in other words, longer fixation time~\cite{just1980theory,clifton2007eye}. Mostly related to our work, ~\cite{gaze-text_garain_2017} and~\cite{unknown-word_hiraoka_2016} leverage features from the gaze and document context to detect difficult words in reading. However, the above methods are all based on eye movement features, such as the number of fixation and regression, so dedicated eye tracking equipment is required to obtain high-precision eye movement data.

The webcam is a ubiquitous device that can track users' eye movements. Researchers apply calibration-free webcam-based eye tracking to predict reading comprehension and mind wandering in reading and achieve comparable accuracy to infrared-based eye tracking on these tasks~\cite{eval-webcam_hutt_2022}. Previous work also utilizes fixation points computed from webcam data to analyze search behaviors, and the mean error of the real-time eye tracking is 128.9 pixels~\cite{searchGazer_papoutsaki_2017}. Although webcam-based eye tracking shows the potential of modeling user behavior, the tracking precision is not enough for detecting unknown words, which requires precise locations of the gaze. 

Based on the strong connection between eye movement and attention in reading, many works introduce gaze to assist model training in Natural Language Processing (NLP) area~\cite{sentiment_long_21,barrett-etal-2018-sequence,ner_Hollenstein_2019}. Since pre-trained language models~\cite{devlin2019bert,liu2019roberta} contain rich text information for downstream tasks, we leverage them to compensate for the inaccuracy of the webcam-based eye-tracking method. By combining gaze and text information, we can accurately detect unknown words for users based on a webcam, which can enable various interactions assisting ESL reading.

\section{GazeReader}
\label{sec:method}

\projectname{} automatically detects unknown words by embedding the gaze data and text information using LSTM and a transformer-based model. We also introduced knowledge-based features including term frequency, part of speech, and named entity recognition to improve the performance. In data pre-processing, we apply a moving average filter and re-sampling to de-noise gaze data and map the gaze data to the text data.

\subsection{Data Preparation}

\begin{figure*}
  \centering
  \includegraphics[scale = 0.45]{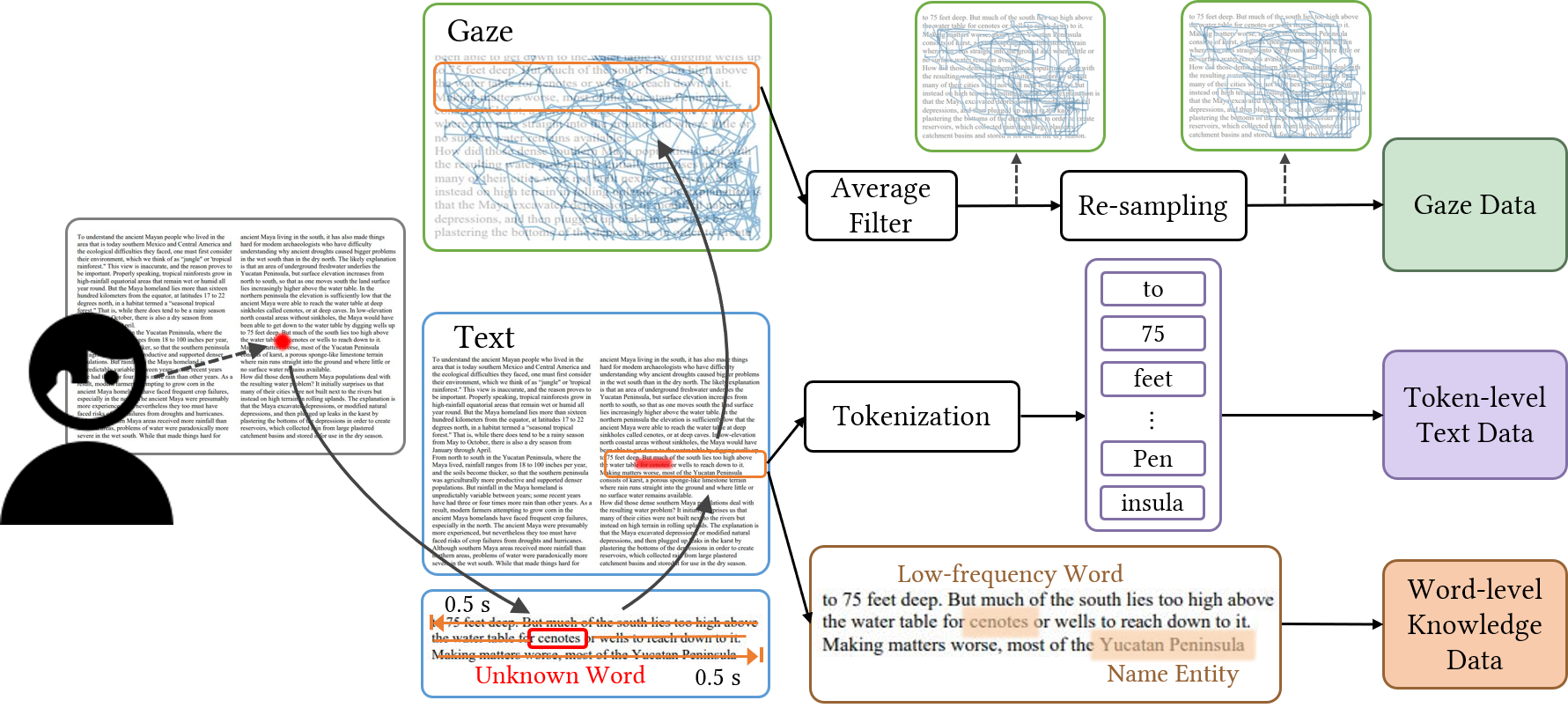}
  \caption{Data pre-processing for \projectname{} model.}
  \label{fig:data_processing} 
\end{figure*}

\subsubsection{Data Collection}
We recruited 12 graduate students ranging in age from 23 to 28 years (\textit{M} = 23.64, \textit{SD} = 1.57), including two males and ten females. They were all second-language learners of English and were able to use English in academic settings, and their years of formal English study ranged from 7 to 24 years (\textit{M} = 16, \textit{SD} = 4.84). Nine of them wore glasses, and the other three did not.
 
We chose 36 articles from TOEFL reading materials with an average length of 386 words per article and 13902 words in total. Participants read 12 articles per day for three days and spent four hours reading in total. The experiment was conducted in a quiet meeting room without disturbing using a Thinkpad X1 carbon laptop (CPU: i7-10710U, 6 cores, 1.1 GHz, RAM: 16GB, Storage: 512GB). There was a break every 30 minutes.

We built a web PDF reader to display the article and used WebGazer.js~\cite{papoutsaki2016webgazer} to track users' gaze. Before each section started, the participants calibrated the WebGazer according to its instruction~\cite{papoutsaki2016webgazer} to map gaze points to screen coordinates. Participants first read the article, recorded the gaze data, and then clicked to mark the unknown words. Our system saved the gaze data, the unknown words, and the text coordinate for each article. We collected a total of 4274 clicks for unknown words and 1232 unknown words after eliminating duplication.

\subsubsection{Data Pre-processing}
Since the gaze data is acquired with only a webcam without highly accurate devices, the original data is extremely noisy and could not be directly fed to the machine learning models. Meanwhile, to avoid potential distractions to users' reading, we do not ask the users to label their unknown words during their reading in real-time, which makes it a necessary step to align the gaze data with its corresponding context and unknown words.

As shown in Fig.~\ref{fig:data_processing}, we first de-noise the gaze data by applying a moving average technique with a sliding window size of 50. Since each time step of the output data from WebGazer is uneven, we re-sample the gaze data with the sample rate of 20Hz and conduct linear interpolation between the raw data on both the x and y-axis. Afterward, to align each unknown word with its corresponding gaze data, we anticipate each word's time based on its relative position of the whole document assuming that people's reading speeds are likely to be uniform. We also chunk the gaze data into segments based on each piece of text's length for matching gaze and text data. We select the text read within 1 second by the user when they see the unknown word as its context. Finally, since we only want to prompt users with the unknown words they are reading, we add negative samples for each gaze data by combining it with the contents containing unknown words that the user was not reading.


\subsection{Unknown Words Detection Model}
\begin{figure*}
  \centering
  \includegraphics[scale = 0.43]{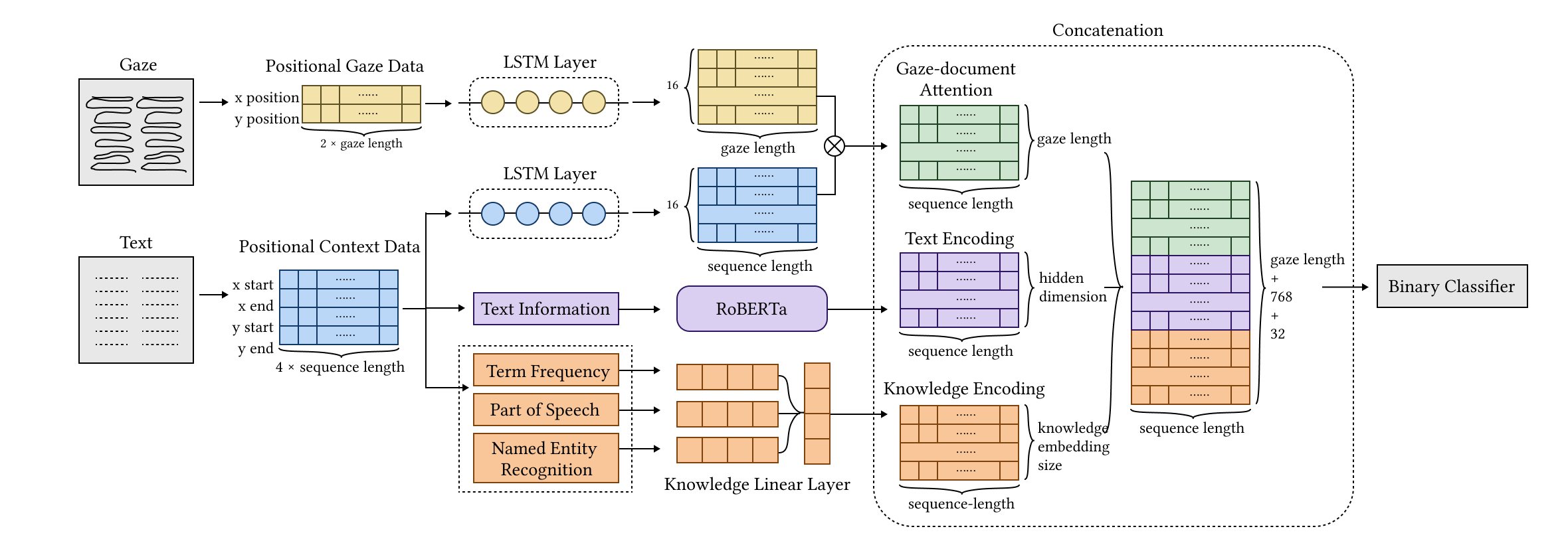}
  \caption{Overview of \projectname{} model.}
  \label{fig:overview} 
\end{figure*}

Our model is designed to predict users' unknown words during reading based on gaze data and reading text information. To capture the positional information of texts and gaze data, we use LSTM layers to encode them into vector space. To improve the model's performance, we use RoBERTa~\cite{liu2019roberta} as the backbone language model for infusing text information. Meanwhile, we further enhance the text information by leveraging word-level knowledge. The overall architecture of our model is illustrated in Fig.~\ref{fig:overview}.

\subsubsection{Positional Data Encoding}
At first, we illustrate our neural network that encodes the sequential 2-D data, including the word position in the reading document and the user's eye gaze data. Since both word position and eye gaze data are sequential, we use LSTM network to encode them since it can capture the long and short-term dependencies in sequential data. Overall, the two LSTM layers transform the 2-D gaze sequence $g \in \mathbb{R}^{2 \times n_{gaze}}$ and word position sequence $d \in \mathbb{R}^{2 \times n_{txt}}$ into $H_g \in \mathbb{R}^{n_p \times n_{gaze}}$ and $H_d \in \mathbb{R}^{n_p \times n_{txt}}$ (where $n_p$ is the output dimension of LSTM layer, while $n_{gaze}$ and $n_{txt}$ are the length of the gaze data and texts), respectively. Afterward, we conduct a matrix multiplication between the encoded text positional data and the gaze data as the attention between the gaze and texts:

\begin{align*}
    g = [g_x;g_y]&, H_g = \textrm{LSTM}(g) \\
    d = [d_x; d_y]&, H_d = \textrm{LSTM}(d) \\
    A_p &= \delta(H_d^T \cdot H_g)
\end{align*}

where $\delta$ is the activation function and $g_x, g_y, d_x, d_y$ are the gaze data and word position on X and Y axis, respectively.

\subsubsection{Context Information Capturing}
In order to leverage the rich information from the reading materials, we use a pre-trained RoBERTa model to encode these texts in the vector space. RoBERTa is a pre-trained language model with 12 transformer layers, where each layer contains a self-attention module that captures the attention between word tokens and a feed-forward layer that maps the attention into a higher vector space. It transforms the input texts $s \in \mathbb{R}^{n_{txt}}$ into $Z \in \mathbb{R}^{n_{txt} \times dim}$, where $dim$ is the hidden dimension of RoBERTa. Since the model has been trained on huge amounts of text data, it can accurately embed documents into vector space for downstream tasks. In general, RoBERTa calculates

\begin{equation}
Z = \textrm{RoBERTa}(s)
\end{equation}

\subsubsection{Knowledge Enhancement}
During the processing of language models, the original documents are tokenized into sub-word tokens instead of words. Therefore, there exists word-level knowledge lost during tokenization which causes the degradation of model performance. Hence, for each token, we add its original word-level knowledge, including term frequency, part of speech~\cite{bird2009natural}, named entity recognition~\cite{honnibal2020spacy} to promote the model's performance. In general, we use learnable embeddings and layers to encode these features into a ``knowledge matrix'' $K \in \mathbb{R}^{n_{txt} \times n_k}$. 

\subsubsection{Training}
We combine these features above and feed them to a linear classifier for unknown word prediction:
\begin{align*}
    O &= [A_p; Z; K] \\
    a &= \sigma(W_o \cdot O+ b_o)
\end{align*}
where $W_o, b_o$ are learnable parameters, $\sigma$ is the sigmoid function, and $a \in \mathbb{R}^{n_{txt}}$ is the output activation of the classifier. The model is trained with a binary entropy loss on all tokens
\begin{equation}
    \mathcal{L} = -\sum_{i=1}^{n_{txt}} (1-\tilde{y}_i) \cdot \log(1-a_i) + \tilde{y}_i \cdot \log(a_i)
\end{equation}
where $\tilde{y}_i$ is the label whether the i-th token is part of the unknown word of the user.

\subsection{User Study}

To demonstrate the variability of our approach, we used the Vocabulary Levels Test (VLT) to test users' vocabulary levels. The test was developed by Nation in 1987 and produced questions at different word frequency levels in 2000, 3000, 5000, 10,000, and academic groups~\cite{nation1990teaching}. 
The VLT is one of the most widely used tests to measure the vocabulary level of English learners ~\cite{read1988measuring,read2000assessing}. We chose an optimized version of the test~\cite{schmitt2001developing}.
Considering that the required vocabulary level for TOEFL reading as experimental material is about 4500~\cite{chujo2009many}, we used a 5000-word frequency level group to test users' vocabulary levels.

We also explored future directions for the tool design. In order to understand users' needs, we designed a group of questions to test their subjective feelings and attitudes. The questions include perceived factors that influence English text reading, behaviors when checking unknown words in reading, concerns toward eye-tracking tools, and attitudes toward our proposed new features (eye-tracking reading assistance and vocabulary management). All the above questions and potential answers were grouped and conducted using 5-point Likert items. 

Besides subjective ratings, we collected descriptive information from randomly selected users about their needs during reading. Users were asked to write down their potential needs as much as possible.

\section{Results and Findings}
\label{sec:results}

\subsection{Prediction Accuracy}

\begin{table}[b]
\begin{tabular}{lrr|rrr}
\hline
PLM Backbone & $n_p$ & $n_k$ & Precision & Recall & F-1 Score \\ \hline
RoBERTa      & 16             & 16           &68.31     &79.20    &73.97           \\
RoBERTa      & 16             & 32                  &\textbf{71.21}           &80.70        &\textbf{75.73}           \\
RoBERTa      & 32             & 16                  &67.29           &84.81        &75.11           \\
RoBERTa      & 32             & 32                  &65.50           &\textbf{86.55}        &74.97           \\ \hline
\end{tabular}
\caption{Unknown word detection results of \projectname.}
\label{tab:main-results}
\end{table}

The main results of our model are shown in Table~\ref{tab:main-results}. We use precision, recall, and F-1 score as metrics for our unknown word detection task. We trained our models for 5 epochs, with the learning rate of 8e-5 for parameters in the RoBERTa and linear layers, and 0.1 for LSTM layers. Our best model can achieve the accuracy of 98.09\% with 75.73\% F-1 score. Moreover, we tried out different hidden dimensions for gaze-text attention $n_p$ and knowledge embedding $n_k$. It shows that by increasing the knowledge dimension $n_k$, the overall performance of our model can be improved while increasing the gaze-document attention dimension $n_p$ helps promote the recall.

\subsection{Transfer Learning Analysis}

\subsubsection{Transfer Learning Setting}
Table~\ref{tab:transfer-learning} indicates our model's transferability between users and reading materials. For the cross-user transfer learning experiment, we trained the model 12 times in total, and for each running, we selected one user's data as the test set and took other users' data as the training set. For the cross-document experiment, we also conducted 12 pieces of training and evaluations. Under each running, we selected the data generated from three documents as the test set, and the rest of the data were combined as the training set. Therefore, our relative data sizes between training and testing data are similar under cross-user and cross-document settings.

\begin{figure}
    \centering
    \includegraphics[width=\columnwidth]{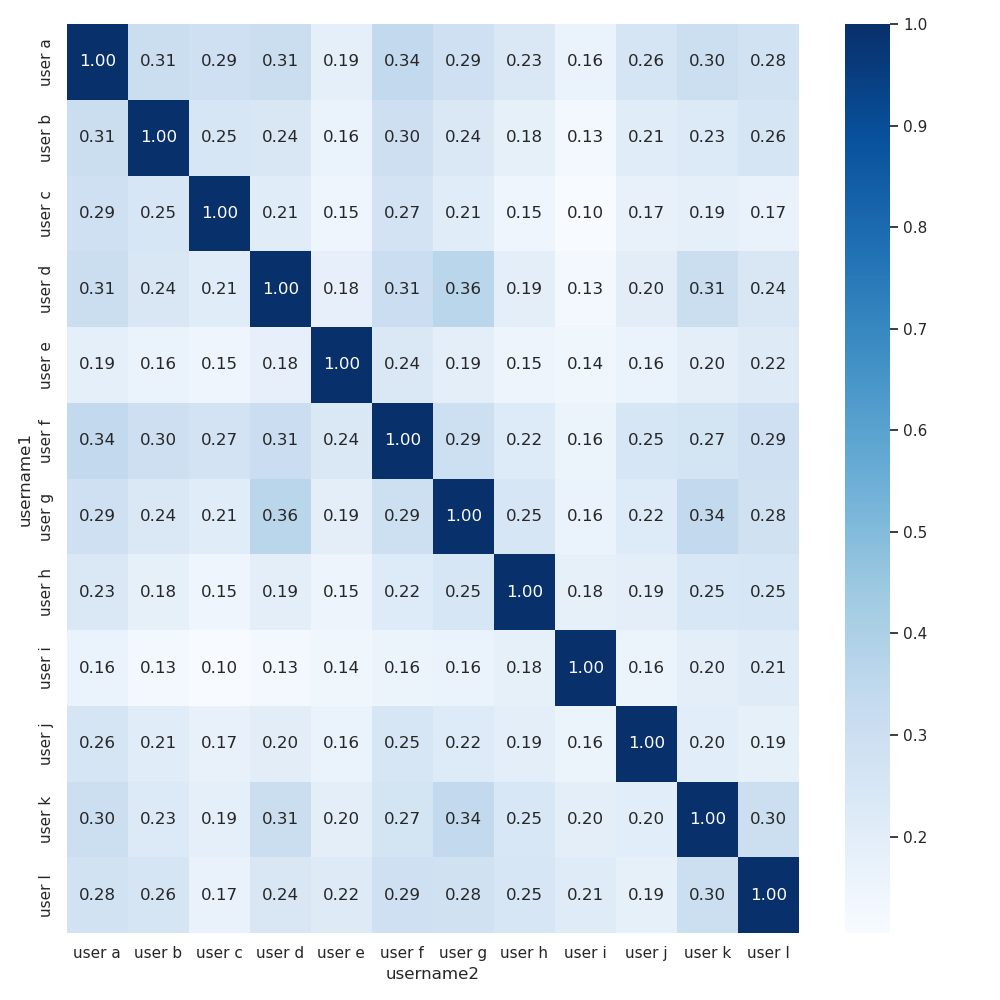}
    \caption{Jaccard similarity of unknown words between users.}
    \label{fig:jaccard}
\end{figure}

\subsubsection{Cross-User Transferability}
We find out that when the model is applied to new users' data, its performance is still on par with its original setting. Therefore, our model can be effectively adapted to new user groups without specific calibrations or tuning, which successfully fits our goal of ubiquitous unknown word detection. We further analyze the unknown words of different users, and their Jaccard similarity matrix is shown in Fig~\ref{fig:jaccard}. The average of cross-user unknown words Jaccard similarity score is 0.23 and most of their similarities are below 0.3, meaning that different users' unknown words are distinct even given the same article. It further proves that our model does not simply remember the unknown words of users, but predicts the unknown words based on the user's behaviors. Meanwhile, the model's performance can become significantly better on some users, and a possible reason is that these users' behaviors are more general, i.e., their language skills might be the average among all people.

\begin{table*}[]
\begin{tabular}{ll|rrr}
\hline
User  dependency & Document dependency & Precision & Recall & F-1 Score \\ \hline
Dependent        & Dependent           &71.21           &80.70        &75.73           \\
Independent      & Dependent           &$73.65 \pm 5.83$           &$82.77\pm 4.41$       &$78.26\pm 4.53$           \\
Dependent        & Independent         &$46.56\pm 4.26$           &$68.39\pm 5.80$        &$56.31 \pm 3.38$           \\ \hline
\end{tabular}
\caption{Transferability analysis of unknown word detection by \projectname{}.}
\label{tab:transfer-learning}
\end{table*}

\subsubsection{Cross-Document Transferability}
In order to test whether our model only memorizes the difficult words from the text, we conducted the experiment to see how the model would perform if the words in the test data were not presented for training. We find that the model's performance dropped in this setting, while the recall is still acceptable. 

\subsection{Ablation Study}

\begin{table*}[]
\begin{tabular}{l|rrr}
\hline
Model                         & Precision & Recall & F-1 Score \\ \hline
\projectname{}                           &\textbf{71.21}           &80.70        &\textbf{75.73}           \\
\projectname{}~w/o context-awareness     &4.78           &59.03        &10.00           \\
\projectname{}~w/o gaze                  &66.35           &\textbf{86.67}        &75.59           \\
\projectname{}~w/o knowledge enhancement &69.96          &80.32        &74.93           \\ \hline
\end{tabular}
\caption{Ablation study of \projectname{}.}
\label{tab:ablation-study}
\end{table*}

We tried to exclude three different modalities of data presented in our model respectively, which are text information, gazing data, and word-level knowledge. After the text information is removed, the model's performance drops significantly, which proves that the use of a pre-trained language model can effectively promote the model's ability to detect unknown words of users. 
Then we removed the gaze data, in which case the model cannot obtain the gaze behavior related to the text. The model's precision drops and the recall increases. It indicates that both the true positive and the false positive increase. In other words, without gaze, the model tends to predict a word as an unknown word. The gaze may help identify whether a difficult word is an unknown word or not.
Finally, the result of ablating knowledge enhancement proves that infusing word-level knowledge can promote the model's performance.

\subsection{Design Guidance}

Our results of the 5000-level VLT show that users can be divided into a high-level group and a low-level group. A Mann-Whitney U test was run to determine if there were differences in the test score between them. Scores for the high-level group (mean rank = 9.5) and low-level group (mean rank = 3.5) were statistically significantly different, \textit{U} = .000, z = -2.887, \textit{p} = .002, using an exact sampling distribution for U. The results demonstrate that our users covered learners of different levels.

As for factors affecting the reading of academic texts, compared to a neutral attitude score of 3, the scores were 3.67 for new words (\textit{SD} = 0.98, \textit{t}(11) = 2.345, \textit{p} < .05), 3.83 for the ambiguous meaning of words (\textit{SD} = 0.72, \textit{t}(11) = 4.022, \textit{p} < .01), 4.25 for long and complex sentences (\textit{SD} = 0.87, \textit{t}(11) = 5.000, \textit{p} < .001), all of which were perceived by users as having a significant effect on comprehension of the text. This result suggests the importance of contextualizing the interpretation of unknown words in reading.

The scores of the two proposed features are 4.35 for eye-tracking reading assistance (\textit{SD} = 0.64, \textit{t}(11) = 7.244, \textit{p} < .001) and 3.93 for vocabulary management (\textit{SD} = 0.99, \textit{t}(11) = 3.260, \textit{p} < .01), respectively. Both are significantly higher than a neutral score of 3, demonstrating that users hold a positive attitude toward our proposed features. The results show that none of the privacy and power consumption is rated significantly, which indicates the possibility of future tool design.

By analyzing words users wrote about expecting features, we found that the three most frequently mentioned features are translating proper nouns, translating long and complex sentences, and accurately identifying multi-meaning words. Two other users also mentioned the need to avoid pop-ups from affecting their reading. We believe these goals can all be achieved using our webcam-based tool and NLP model.

Based on the results mentioned above, applying the webcam to unknown words detecting and reading assistance by our approach of eye tracking and NLP model has theoretical and practical value. Interaction design strategies should also be addressed in developing applications to improve the user experience.
\section{Discussion and future work}
\label{sec:discussion}

The accuracy of our model is high by combining the information from the gaze, text, and knowledge, but the F1-score is only 75.73\%. The reason could be that unknown words are sparse compared with other words. It causes an unbalance in the dataset. In addition, the size of the dataset is relatively small compared with the parameter in the neural network might be the other reason. More case studies are needed to find out the reason.

The ablation shows that the F1-score doesn't significantly decrease when we remove the gaze from the model. One of the potential reasons is the inaccuracy of the webcam-based eye-tracking algorithm we used. WebGazer~\cite{searchGazer_papoutsaki_2017} uses the coordinates of clicks to calibrate the position of gaze points, so tracking accuracy becomes worse when participants read without using a mouse. The optimization of the eye-tracking algorithm is needed to make gaze contribute more to the performance. Multi-modal data obtained by earphones and smart glasses can be used to infer head orientation and position~\cite{faceori,reflectrack} to facilitate the eye-tracking algorithm.

Our data is collected on a commodity laptop using an eye-tracking library that can work on any website~\cite{papoutsaki2016webgazer}. Therefore, our method should be able to generalize to other models of computers, though we only used one type of laptop to collect data in this paper. In addition to the current web-based implementation, the same algorithm should be able to migrate to various applications.

We took TOEFL reading passages as reading materials in the experiment, considering that unknown words in academic text create more difficulty in reading. The TOEFL readings cover a variety of topics, so our method has the potential to work on different types of articles. In the future, we will use more diverse materials to make our approach more generalizable. Participants in our experiment are all graduate students in the United States. We regard them as our target users because they have a more urgent need to read more efficiently. Our participants consisted of learners with different vocabulary levels, and we will also include more users of different learning stages in the future.

Applying our approach to languages other than English is another aspect of future work. We believe that this work may be generalized to other alphabetic languages because alphabetic languages have similar word structures and some of them even have the same alphabet. It remains to be seen whether gaze data can be used to detect unknown words in non-alphabetic languages.

For applications, future work could explore how to provide proper assistance to second language users during reading according to our exploratory user study. We aim to design a simple webcam-based tool across different levels of users that has high efficiency and usability at the same time. Exploring ways to properly present the information that users need would be a compelling future study.

\section{Conclusion}
\label{sec:conclusion}

We propose \projectname{}, a novel unknown word detecting method using the webcam for ESL learners. \projectname{} leverages the gaze data obtained from the webcam and introduces the transformer-based model to enhance the performance. \projectname{} can detect unknown words with the accuracy of 98.09\% and the F1-score of 75.73\%. We also conducted a user study to explore the design scope based on \projectname{}. The result shows that the design focus should be proper nouns, multi-meaning words, and long and complex sentences. Users show positive attitudes towards the design features based on unknown word detection. As the webcam becomes more ubiquitous with the prevalence of online working and studying, \projectname{} can enable a wide range of applications and benefit a large population of ESL learners in reading.

\begin{acks}
This work is supported by the Natural Science Foundation of China (NSFC) under Grant  No. 62132010 and No. 62002198, Young Elite Scientists Sponsorship Program by CAST 
under Grant No. 2021QNRC001, Tsinghua University Initiative Scientific Research Program, Beijing Key Lab of Networked Multimedia, Institute for Artificial Intelligence, Tsinghua University.
\end{acks}

\bibliographystyle{ACM-Reference-Format}
\bibliography{ref}

\end{document}